\documentclass[epj]{svjour}
\usepackage{graphicx}
\begin{document}
\title{Characterization of Bernstein modes in quantum dots}
\titlerunning{Bernstein modes in quantum dots\ldots}
\author{Manuel Val\'{\i}n-Rodr\'{\i}guez \inst{1}
\thanks{E-mail: \email{VDFSMVR4@clust.uib.es}}
Antonio Puente \inst{1}, 
Lloren\c{c} Serra \inst{1},
Vidar Gudmundsson \inst{2} 
\and Andrei Manolescu \inst{2}}
\authorrunning{M. Val\'{\i}n-Rodr\'{\i}guez {\em et al.}}
\institute{
Departament de F\'{\i}sica,
Universitat de les Illes Balears, E-07071 Palma de Mallorca, Spain
\and
Science Institute, University of Iceland, Dunhaga 3, 
IS-107 Reykjavik, Iceland}
\date{October 15, 2001}
\abstract{
The dipole modes of non-parabolic quantum dots are studied
by means of their current and density patterns 
as well as with their local absorption distribution.
The anticrossing of the so-called Bernstein modes originates from
the coupling with electron-hole excitations of the two
Landau bands which are occupied at the corresponding magnetic 
fields.
Non-quadratic terms in the potential cause an energy separation between 
bulk and edge current modes in the anticrossing region. 
On a local scale the 
fragmented peaks absorb energy in complementary spatial regions
which evolve with the magnetic field.
} 

\maketitle

\section{Introduction}

The far infrared (FIR) spectroscopy has proved to be an invaluable tool
for the physical characterization of semiconductor quantum dots
and other electronic nanostructures \cite{Sik89,Dem91}. 
Since its initial applications
to 2D semiconductor quantum dots one of the main motivations has been 
to identify a signal of the relative motion of the confined electrons.
As is well known \cite{boo1}, most commonly the confinement seems 
to be parabolic, and in this limit the generalized Kohn theorem 
assures that the only allowed excitations are the 
center-of-mass modes, at the frequencies
\begin{equation}
\label{eq1}
\omega_\pm(B)=\sqrt{\omega_0^2+\frac{\omega_c^2}{4}}
\pm
\frac{\omega_c}{2}\; ,
\end{equation}   
where $B$ is the applied perpendicular magnetic field, $\omega_0$ is 
the frequency associated with the external parabola
and $\omega_c=eB/c$ is the cyclotron frequency \cite{units0}.

In Ref.\ \cite{Gud95} the fragmentation of the high-energy branch
$\omega_+(B)$ was measured in wires and dots and interpreted as an 
interaction with the cyclotron harmonics at energies 
$n\omega_c$ ($n=2,3$). Since these interactions resemble the 
Bernstein modes of the electron gas \cite{Ber58}, the same name 
was used to label the corresponding excitations in these nanostructures. 
More recently,
Krahne {\em et al.} \cite{Kra00} have measured the Bernstein modes
in a 2D GaAs system with tunable electron density, varying from a 
continuous 2D gas to well separated dots. While the anticrossing exactly 
occurs at $2\omega_c$ for the 2D gas, for finite quantum dots it lies 
below (between $\omega_c$ and $2\omega_c$). 

It is our aim in this paper
to provide a physical characterization of the below-$2\omega_c$ Bernstein
modes of quantum dots by analysing the current and density distributions,
as well as the local absorption patterns, associated
with each particular peak of the dipole spectrum.  
For this purpose, we shall use the time-dependent local-spin-density 
approximation in a symmetry unrestricted formalism
as developed recently in Ref.\ \cite{Pue99}. A comparison with the 
electron-hole interband transitions will also shed light on the 
origin of the fragmentation of the $\omega_+(B)$ branch.

\section{Theoretical approach}
As in Ref.\ \cite{Gud95} we assume the following confining potential
\begin{equation}
\label{eq2}
v_{\rm conf}(r)=\frac{1}{2}\omega_0
\left[
\left(\frac{r}{\ell_0}\right)^2+
a \left(\frac{r}{\ell_0}\right)^4
\right] \; ,
\end{equation}
with $r$ the radial coordinate in the $xy$ plane, where the 
electronic motion occurs, $\ell_0=\omega_0^{-1/2}$
the {\em confining length},
$\omega_0=3.37$~meV and $a=2.02\;10^{-2}$. 
The parameter $a$ controls the degree
of non-parabolicity although, as shown in Ref.\ \cite{Gud95},
the Bernstein fragmentation discussed below 
is actually not much sensitive on its precise value.
It is worth to mention that a similar potential including a 
quartic term was used by Ye and Zaremba to analyze the breaking of 
Kohn's theorem in the context of a hydrodynamic approach
\cite{Ye94}.

We assume the effective-mass Hamiltonian, applicable to GaAs 
nanostructures \cite{Ham}, and describe the electronic exchange and
correlation effects within the Local-Spin-Density approximation (LSDA).
This density-functional approach has recently been used by several 
authors to describe quantum dot properties. The reader is addressed 
to Refs.\ \cite{Fer94,Kos99,Hir99} for details on the method. 
In the context of the FIR absorption the time-dependent extension 
(TDLSDA) has also been shown to provide adequate 
results \cite{Ull00,Pue01}.

The dynamical properties described below have been calculated
by integrating the Kohn-Sham equations
\begin{equation}
\label{eqH}
i\frac{\partial\varphi_{i\eta}({\bf r},t)}{\partial t}
= h_\eta[\rho,m]\; \varphi_{i\eta}({\bf r},t)\; ,
\end{equation} 
where $\eta=\uparrow,\downarrow$ labels the two spin components,
while 
$\rho=\sum_{\rm occ.}{|\varphi_{i\uparrow}|^2 
                    + |\varphi_{i\downarrow}|^2}$ 
indicates the total particle density and
$m=\sum_{\rm occ.}{|\varphi_{i\uparrow}|^2-|\varphi_{i\downarrow}|^2}$ 
the total spin magnetization. The self-consistent Hamiltonian in Eq.\
\ref{eqH} reads
\begin{eqnarray}
\label{eqHh}
h_\eta[\rho,m]&=& \frac{1}{2} \,\left(-i\nabla+\frac{e}{c} 
{\bf A}({\bf r})\right)^2 
+v_{\rm conf}(r)\nonumber\\ 
&+& \int{d{\bf r}' {\rho({\bf r}')\over |{\bf r}-{\bf r}'|}}
+ {\delta E_{XC}[\rho,m]\over\delta\rho} \nonumber\\
&+& s_\eta \left( {\delta E_{XC}[\rho,m]\over\delta m}
+ g^* \mu_B \frac{B}{2} \right) 
\; ,
\end{eqnarray}
with ${\bf A}({\bf r})=B/2(-y,x)$ being the vector potential in the 
symmetric gauge, 
$E_{XC}[\rho,m]$ the exchange-correlation functional, and $s_\eta=\pm 1$
for $\eta=\uparrow,\downarrow$, respectively. The Zeeman term in 
Eq.\ \ref{eqHh} contains the gyromagnetic factor $g^*$ and the Bohr
magneton $\mu_B=e/2m_ec$.

\section{Single peak analysis}

After calculating the ground state structure and in order to excite 
the collective dipole mode we perform a small
rigid translation of the electronic cloud. This takes the system out
of equilibrium and as a consequence it begins to oscillate.
The expectation values of several observables ${\cal O}$ are subsequently 
recorded in time $\langle {\cal O}\rangle(t)$ and frequency analysed to
obtain the corresponding energy distributions. 
In the present analysis
we have considered the dipole, local current and local density 
operators, i.e.,
\begin{eqnarray}
\label{ops}
\hat{D}&=&\sum_i{x_i+y_i}\; ,\nonumber\\
\hat{\bf j}({\bf r})&=&\sum_i{
\left[ -\frac{i}{2}
\left(\stackrel{\rightarrow}{\nabla}_i-
\stackrel{\leftarrow}{\nabla}_i\right)
+\frac{e}{c}{\bf A}({\bf r}_i)\right]
\delta({\bf r}-{\bf r}_i)}\; ,
\nonumber\\
\hat{\rho}({\bf r})&=&\sum_i{\delta({\bf r}-{\bf r}_i)}
\; .
\end{eqnarray}  
Note that after the frequency analysis we obtain 
$\langle {\cal O}\rangle(\omega)$ which for the
local signals
provides a pattern of currents 
$\langle \hat{\bf j}({\bf r})\rangle(\omega)$
and of local density $\langle \hat{\rho}({\bf r})\rangle(\omega)$ that
can obviously be ascribed to the excitation at energy $\omega$.
At a fixed $\omega$, the time evolution at every point ${\bf r}$
is simply 
given by a phase $e^{-i\omega t}$, or by $\sin(\omega t)$ and 
$\cos(\omega t)$ for the corresponding real transforms, 
thus permitting to monitor the variation of the
current and density patterns for 
each excitation peak. In addition, the density patterns
can be used to obtain the local absorption from
$|\,\langle \hat{\rho}({\bf r})\rangle(\omega)\,|$,  as done 
in Ref.\ \cite{Val01} for the characterization 
of the FIR absorption of triangular and square quantum dots. 
Theoretically, the local absorption at ${\bf R}$ provides
the energy absorbed by the system when a probe of the 
type $\xi({\bf r}-{\bf R})$ is used, where 
$\xi$ is a highly peaked spatial modulation \cite{Sim00}. 
Note also that in Eq.\ \ref{ops} we have used 
the gauge invariant current including explicitly the vector potential
${\bf A}({\bf r})$.

\section{Results}

Figure 1 displays the FIR absorption for a dot with $N=20$
electrons in the confining potential (\ref{eq2}) for 
different vertical magnetic fields. The Bernstein fragmentation of the 
high energy branch is conspicuous for $1.6\;{\rm T}<B<2.4\;{\rm T}$. 
Actually, for several $B$'s up to three peaks can be identified 
in the higher branch, 
although in most instances this branch fragments into two dominant 
features.
In this figure the dashed line indicates the $2\omega_c$ values while 
the dotted lines
show the particle-hole transitions with $\Delta\ell=+1$, which by 
angular-momentum selection rules are the only ones that couple
to the high-energy branch. 
All these transitions are of
interband character. The $\Delta\ell=-1$ particle-hole transitions 
(of both intra- and interband character)
only contribute to the low-energy branch and thus are not 
relevant for the present discussion.
From this figure and the results that will be presented below,
the fragmentation of the FIR absorption, causing the Bernstein modes, 
can be 
understood as an effect of the coupling with particle-hole
transitions, allowed by the non-parabolicity of the confining 
potential. Note that at $B\approx 2$~T there is a 
particle-hole excitation actually lying in between the fragmented
Bernstein peaks, thus showing that the maximal 
overlap of the interacting response
with the interband transitions occurs below 
the $2\omega_c$ line. We stress that in spite of the overlap
with $\omega_-(B)$, the displayed transitions do not
couple with the lower branch by angular momentum selection rules.

The relevance of the different particle-hole transitions is more
easily appreciated within the perturbative response formalism, as 
opposed to the real-time one (see, e.g., Ref.\ \cite{long}). 
Using the perturbative method, in Fig.\ 2 we 
compare the full absorption function at $B=1.8$ T with those obtained by
including only the hole states from the first and second Landau bands,
respectively, shown in the left panel. The interacting responses in 
the restricted subspaces show collective peaks which approximately 
reproduce the energies of the Bernstein excitations in the full response.
Therefore, each of the Bernstein peaks can be thought of as arising 
from particle-hole excitations of different bands, as it happens
in the bulk limit for a modulated 2D gas \cite{Man98}.
Note, however, that the picture of separate FIR excitations for each
band is an approximate one, since there are important interference 
effects between bands. 
With this interpretation we expect that when 
$\omega_{+}(B)$ and $2\omega_c(B)$ intersect in a region where only
one Landau band is occupied there will be no Bernstein fragmentation 
as in Fig.\ 1. Indeed we have checked this by using a much lower value 
of the confinement, $\omega_0=0.5$~meV, which for $B=0.3$~T
has one (spin degenerate) occuppied Landau band and  
$\omega_{+}(B)\approx 2\omega_c(B)$ without any anticrossing.

The preceding analysis is in qualitative agreement with 
the results for Raman modes in quantum wires of Steinebach
{\em et al.} \cite{Ste96}. These authors attribute the anticrossing 
to the large importance of transitions with $\Delta n=2$, where 
$n$ is the Landau level index, with respect to those with
$\Delta n=1$. 
Realizing that $\Delta n=1$ transitions are mainly  
from the highest occupied Landau band, while
$\Delta n=2$ also includes the second highest Landau band,
one arrives at a similar conclusion as the above one.     

We consider next the current distributions.
The pure Kohn modes in parabolic dots correspond to rigid translations 
of the electronic density and, therefore, to essentially uniform
current distributions, the only variations being due to the density
inhomogeneities. A similar 
result is found for the low energy branch in Fig.\ 1, indicating 
the Kohn-like character of this mode.
 
The current patterns at a given time for the Bernstein modes
in the nonparabolic dot are shown in Fig.\ 3.
The time evolution for each of these patterns is simply an 
anticlockwise rotation, in agreement with the $\Delta\ell=+1$ character
of the mode (the positive $z$ axis is pointing towards the reader). 
The leftmost panel 
corresponds to the dominant peak at $B=1.4$~T, which
displays a current pattern basically uniform in the bulk of the dot, thus 
similar to the previously mentioned Kohn modes.
The two intermediate panels correspond to the 
peaks at $B=2$~T (at $\omega\approx0.52$ and 0.54). They show an 
incipient separation of bulk and edge current oscillations which is further
developed at $B=2.4$~T (two right panels). Note that the edge-current
patterns contain a hole in the dot center. 
Therefore, the non-parabolicity of the  
potential induces a separation in energy of the higher-branch 
bulk and edge current modes, which are degenerate in the purely 
parabolic case.

It is also interesting to look at the density variation patterns 
displayed in Fig.\ 4 for the $B=2.4$~T case. As before, the 
time evolution of each pattern is a clockwise rotation for the 
lower-energy branch and anticlokwise for the upper modes. 
Red and blue indicate increment and decrement in the 
local density, respectively. Note that both the low-energy peak (f1)
and the dominant peak of the higher branch (f2) exhibit a simple 
dipole pattern with two lobes. On the contrary, the highest mode (f3)
displays an internal structure reflecting the existence of a node
in the radial density and four regions with alternate phases.
The internal structure of f3 nicely correlates with the the current 
pattern of Fig.\ 3, since the regions of current convergence (divergence)
correspond to an increase (decrease) of the local density.
 
A time average of the oscillating density amplitudes highlights
the regions of higher FIR absorption in the dot, as shown in Fig.\ 5.  
The different grey colours indicate the 
absorption strength as given by the local oscillation.
We notice that at $B=1.4$~T the Bernstein peaks 
absorb basically in rings, with the lower 
peak having a more internal character. As the magnetic field
is raised the lower peak tends to expand its absorption ring,
while the higher one absorbs more in the inner region.
As a result, at $B\approx 2.6$~T the two modes reverse their
character, the upper mode becoming more internal. Thus showing 
that the two Bernstein peaks absorb energy in complemementary 
spatial regions.

\section{Conclusions}

We have analysed the fragmentation of the high-energy 
branch in non-parabolic dots by using a) a comparison with 
the allowed electron-hole transitions, b) the current and density 
variations associated with each peak, and c) the local absorption pattern
of each mode. We conclude that the Bernstein fragmentation 
is a result of the coupling with electron-hole dipole transitions
originating from each of the two occupied Landau bands. 
This coupling manifests in an energy separation 
of the bulk and edge current modes, with the latter one having 
a four-lobe structure in the oscillating density.
The separation of the modes is also reflected
in the local absorption, being more internal for the 
lower peak than for the higher one at small $B$'s and reverting character 
when the magnetic field is increased.


This work was supported by Grant No.\ PB98-0124 from DGESeiC, 
Spain, the Research Fund of the University of Iceland, and the Icelandic
Natural Science Council.

\begin{figure*}[f]
\centerline{\includegraphics*[width=4.in,clip=]{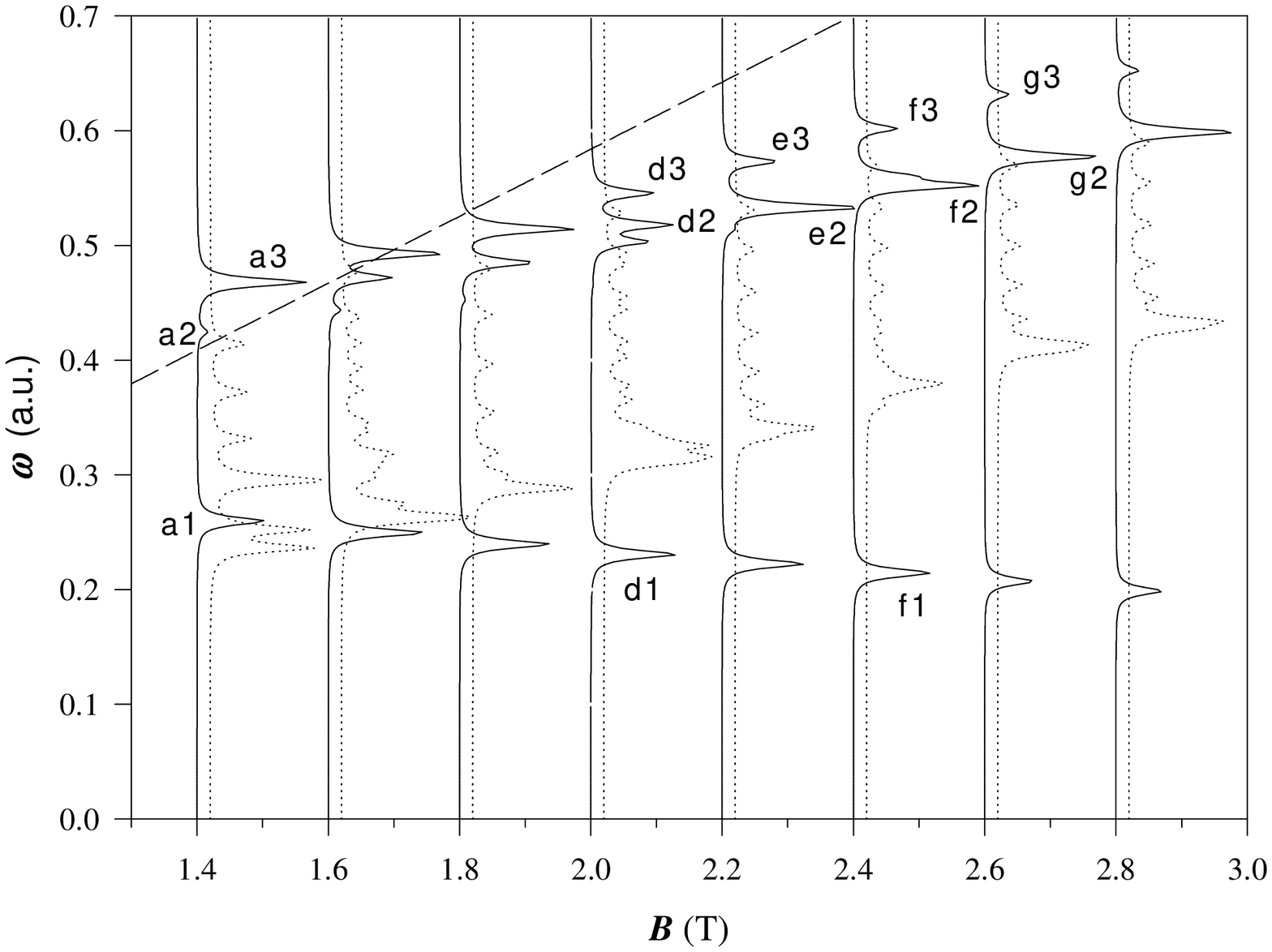}}
\caption{Evolution of the dipole strength for the 20-electron dot.
The dashed line indicates $\omega=2\omega_c$ and the dotted one the 
$\Delta\ell=+1$
electron-hole interband transitions. The labels identify the different 
peaks discussed in the text.} 
\label{Fig1}
\end{figure*}

\begin{figure*}[f]
\centerline{\includegraphics*[width=4.in,clip=]{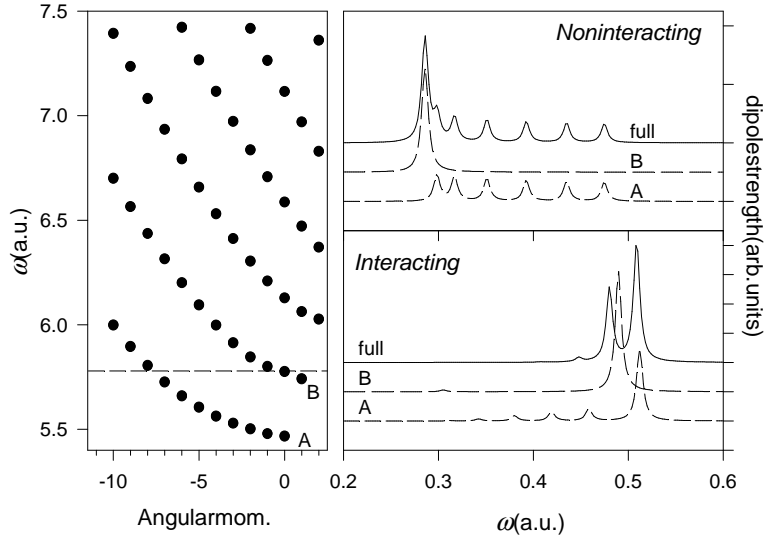}}
\caption{Left panel displays the Kohn-Sham single particle energies
as a function of angular momentum at a magnetic field $B=1.8$~T.
The dashed line separates occupied and non-occupied levels.
Right panels show the response using the perturbative 
approach \cite{long}.
The dashed lines correspond to the restricted subspaces for hole states
in Landau bands A and B, respectively.} 
\label{Fig1b}
\end{figure*}

\begin{figure*}[f]
\centerline{\includegraphics*[width=5.1in,clip=]{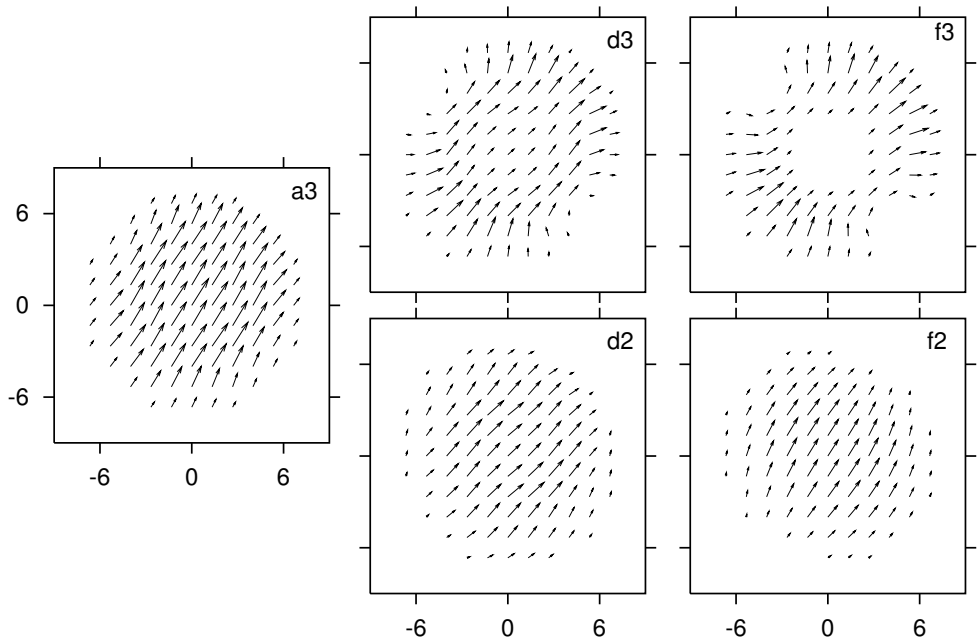}}
\caption{Current density patterns for the high energy branch 
at different magnetic fields. Each label
indicates the corresponding mode in Fig.\ 1. A different absolute 
scale, varying in proportion to the dipole strength of the 
corresponding peak of Fig.\ 1, has been used in each panel.} 
\label{Fig3}
\end{figure*}

\begin{figure*}[f]
\caption{Pattern of local-density variations for the indicated
modes of Fig.\ 1 which correspond to the three dipole peaks
at $B=2.4$~T.
The gray color scale indicates
the signal magnitude while the corresponding sign is superimposed
in white.   } 
\label{Fig4}
\end{figure*}

\begin{figure*}[f]
\caption{Local absorption patterns for the Bernstein peaks
at different magnetic fields. Black regions correspond to high 
absorption, with an absolute scale proportional to the peak's
dipole strength.}
\label{Fig5}
\end{figure*}


\begin{thebibliography}{}

\bibitem{Sik89} C. Sikorski and U. Merkt,
Phys.\ Rev.\ Lett.\ {\bf 62}, 2164 (1989).

\bibitem{Dem91} T. Demel, D. Heitmann, P. Grambow, and K. Ploog,
Phys.\ Rev.\ Lett.\ {\bf 66}, 2657 (1991).

\bibitem{boo1} {\em Quantum dots}, L. Jacak, P. Hawrylak, and
A. W\'ojs (Springer, 1998). 

\bibitem{units0} We use the effective-atomic-unit system in the 
context of the effective-mass Hamiltonian. In terms of the 
the semiconductor dielectric constant $\kappa$ and electron effective 
mass $m$ we thus impose $e^2/\kappa=m=\hbar^2=1$. 

\bibitem{Gud95} V. Gudmundsson, A. Brataas, P. Grambow, B. Meurer,
T. Kurth, and D. Heitmann, 
Phys.\ Rev.\ B {\bf 51}, 17\,744 (1995).

\bibitem{Ber58} I. B. Bernstein,
Phys.\ Rev.\ {\bf 109}, 10 (1958).

\bibitem{Kra00} R. Krahne, M. Hochgr\"afe, Ch.\ Heyn, and D. Heitmann,
Phys.\ Rev.\ B {\bf 61}, 16\,319 (2000).

\bibitem{Pue99} A. Puente, Ll. Serra, Phys.\ Rev.\
Lett.\ {\bf 83} 3266 (1999).

\bibitem{Ye94} Z. L. Ye and E. Zaremba,
Phys.\ Rev.\ B {\bf 50}, 17217 (1994).

\bibitem{Ham} Taking the GaAs values for the electron 
effective mass $m=0.0667m_e$, dielectric constant $\kappa=12.4$ and 
electron giromagnetic factor $g^*=-0.44$, the associated effective 
atomic units are the modified Hartree $H^*\approx 12$~meV, and effective
Bohr radius $a_0^*\approx 100$~\AA. 

\bibitem{Fer94} M. Ferconi and G. Vignale,
Phys.\ Rev.\ B {\bf 50}, 14722 (1994).

\bibitem{Kos99} M. Koskinen, M. Manninen, S. M. Reimann,
Phys.\ Rev.\ Lett.\ {\bf 79}, 1389 (1999).

\bibitem{Hir99} K. Hirose and N. S. Wingreen,
Phys.\ Rev.\ B {\bf 59}, 4604 (1999).

\bibitem{Ull00} C. A. Ullrich and G. Vignale,
Phys.\ Rev.\ B {\bf 61}, 2729 (2000).

\bibitem{Pue01} A. Puente, Ll.\ Serra, 
Phys.\ Rev.\ B {\bf 63}, 125334 (2001).

\bibitem{Val01}
M. Val\'{\i}n-Rodr\'{\i}guez, A. Puente, and Ll.\ Serra,
Phys.\ Rev.\ B {\bf 64}, 205307 (2001).

\bibitem{Sim00}
C. D. Simserides, U Hohenester, G. Goldoni, and E. Molinari, 
Phys.\ Rev.\ B {\bf 62}, 13657 (2000).

\bibitem{long}
Ll.\ Serra, M. Barranco, A. Emperador, M. Pi, E. Lipparini,
Phys.\ Rev. B {\bf 59}, 15290 (1999).

\bibitem{Man98}
A. Manolescu and V. Gudmundsson,
Phys.\ Rev. B {\bf 57}, 1668 (1998).

\bibitem{Ste96} C. Steinebach, R. Krahne, G. Biese, C. Sch\"uller, 
D. Heitmann, and K. Eberl, 
Phys.\ Rev. B {\bf 54}, 14281 (1996).

\end{thebibliography}
\end{document}